\documentclass{acm_proc_article-sp}
\usepackage{times}

\usepackage{stmaryrd}

\usepackage{amsfonts}
\usepackage{amssymb}
\usepackage{graphicx}
\usepackage{subfigure}
\usepackage{array}
\usepackage{hyperref}

\begin{document}
%
\title{Mining Concurrent Topical Activity in Microblog Streams}
\author{A. Panisson, L. Gauvin, M. Quaggiotto, C. Cattuto\\
 Data Science Laboratory, ISI Foundation, Torino, Italy\\
 {\fontsize{10}{6}\selectfont{\{andre.panisson\},\{laetitia.gauvin\},\{marco.quaggiotto\},\{ciro.cattuto\}@isi.it}}
 }
\maketitle
\begin{abstract}
\small
\begin{quote}
Streams of user-generated content in social media
exhibit patterns of collective attention across diverse topics, with temporal structures determined both by exogenous factors
and endogenous factors. Teasing apart different topics and resolving their individual, concurrent, activity timelines is a key challenge in extracting knowledge from microblog streams. Facing this challenge requires the use of methods that expose latent signals by using term correlations across posts and over time.
Here we focus on content posted to Twitter during the London 2012 Olympics, for which a detailed schedule of events is independently available and can be used for reference. We mine the temporal structure of topical activity by using two methods based on non-negative matrix factorization.
%
%
%
%
%
%
We show that for events in the Olympics schedule that can be semantically matched to Twitter topics, the extracted Twitter activity timeline closely matches the known timeline from the schedule.
Our results show that, given appropriate techniques to detect latent signals, Twitter can be used as a social sensor to extract topical-temporal information on real-world events at high temporal resolution.
%
\end{quote}
\end{abstract}

%


\section{Introduction}
\label{sec:intro}

Streams of user-generated content from social media and microblogging systems exhibit patterns of collective attention 
across diverse topics, with temporal structures determined both by exogenous factors, such as driving from mass media, 
and endogenous factors such as viral propagation. Because of the openness of social media, of the complexity of their 
interactions with other social and information systems, and of the aggregation that typically leads to the observable stream 
of posts, several concurrent signals are usually simultaneously present in the post stream, corresponding to the activity of 
different user communities in the context of several different topics. Making sense of this information stream is an inverse 
problem that requires moving beyond simple frequency counts, towards the capability of teasing apart latent signals that 
involve complex correlations between users, topics and time intervals.

The motivation for the present study is twofold.
On the one hand, we want to devise techniques that can reliably solve the inverse problem of extracting latent signals of 
attention to specific topics based on a stream of posts from a micro-blogging system. That is, we aim at extracting the time-varying topical structure of a microblog stream such as Twitter.
On the other hand, we want to deploy these techniques in a context where temporal and semantic metadata about external 
events driving Twitter are available, so that the relation between exogenous driving and time-varying topical responses can 
be elucidated.
We do not regard this as a validation of the methods we use, because the relation between the external drivers and the 
response of a social system is known to be complex, with memory effects, topical selectivity, and different degrees of 
endogenous social amplification. Rather, we regard the comparison between the time-resolved topical structure of a 
microblog stream and an independently available event schedule as an important step for understanding to what extent 
Twitter can be used as a social sensor to extract high-resolution information on concurrent events happening in the real 
world.

Here we focus on content collected by the Emoto project\footnote{http://www.emoto2012.org} from Twitter during the London 2012 Olympics, for which a daily schedule of the starting 
time and duration of sport events and social events is available and can be used for reference.
In this context, resolving topical activity over time requires to go beyond the analysis and characterization of popularity 
spikes. A given topic driven by external events usually displays an extended temporal structure at the hourly scale, with 
multiple activity spikes or alternating periods of high and low activity.  We aim at extracting signals that consists of an 
association of (i) a weighted set of terms defining the topic, (ii) a set of tweets that are associated to the topic, together with 
the corresponding users, and (iii) an activity profile for the topic over time, which may comprise disjoint time intervals of 
nonzero activity.
We detect time-varying topics by using two independent methods, both based on non-negative matrix or tensor factorization. 
In the first case we build the full tweet-term-time tensor and use non-negative tensor factorization to extract the topics and 
their activity over time. We introduce an adapted factorization technique that can naturally deal with the special tensor 
structure arising from microblog streams.
In the second case, which in principle affords on-line computation, we build tweet-term frequency matrices over 
consecutive time intervals of fixed duration. We apply non-negative matrix factorization to extract topics for each time 
interval and we track similar topics over time by means of agglomerative hierarchical clustering.
%

We then apply both methods to the Twitter dataset collected during the Olympics, which reflects the attention users pay to 
tens of different concurrent events over the course of every day. We focus on topical dynamics at the hourly scale, and find 
that for those sport events in the schedule that can be semantically matched to the topics we obtain from Twitter, the activity 
timeline of the detected topic in Twitter closely matches the event timeline from the schedule.
%


This paper is structured as follows: Section~\ref{sec:background} reviews the literature on collective attention, popularity, 
and topic detection in microblog streams. Section~\ref{sec:data} describes the Olympics 2012 Twitter dataset used for the 
study, the event schedule we use as an external reference, and introduces some notations and conventions used 
throughout the paper. Section~\ref{sec:TF} and Section~\ref{sec:MF} describe the two techniques we use to mine time-varying
topical activity in the Twitter stream.
Section~\ref{sec:results} discusses the relation between the time-varying 
topics we obtain and the known schedule of the Olympics events for one representative day, and provides some general 
observations on the behavior of the two methods. Finally, Section~\ref{sec:conclusions} summarizes our findings
and points to directions for further research.


\section{Related Work}
\label{sec:background}



The dynamics of collective attention and popularity in social media has been the object of extensive investigation in the literature. Attention can suddenly concentrate on a Web page~\cite{wu07,ratkiewicz10b}, a YouTube video~\cite{crane08,figueiredo11,naaman11}, a story in the news media~\cite{leskovec09}, or a topic in Twitter~\cite{kwak10,asur11,yang11}.
Intrinsic features of the popular item under consideration have been related to its popularity profile by means of semantic analysis and natural language processing of user-generated content~\cite{adar07,huang10,wu11}.
In particular, a great deal of research~\cite{crane08,kwak10,laniado10,yang11,Lehmann:2012} has focused on characterizing the shape of peaks in popularity time series and in relating their properties to the popular item under consideration, to the relevant semantics, or to the process driving popularity.

Within the broad context of social media, Twitter has emerged as a paradigmatic system for the vision of a ``social sensor'' that can be used to measure diverse societal processes and responses at scale~\cite{jansen09,Sakaki:2010,Bollen2011,mocanu2012twitter}.
To date, comparatively little work has been devoted to extracting signals that expose complex correlations between topics and temporal behaviors in micro-blogging systems. Given the many factors driving Twitter, and their highly concurrent nature, exposing such a topical-temporal structure may provide important insights in using Twitter as a sensor when the social signals of interest cannot be pinpointed by simply using known terms or hashtags to select the relevant content, or when the topical structure itself, and its temporal evolution, needs to be learned from the data.
Saha and Sindhwani~\cite{Saha:2012} adopt such as viewpoint and propose an algorithm based on non-negative matrix factorization that captures and tracks topics over time, but is evaluated at the daily temporal scale only, against events that mainly consist of single popularity peaks, without concurrency.
Here we aim at capturing multiple concurrent topics and their temporal evolution at the scale of hours, in order to be able to compare the extracted signals with a known schedule for several concurrent events taking place during one day.

As we will discuss in detail, microblog activity can be represented using a tweet-term-time three-way tensor, and tensor factorization techniques can be used to uncover latent structures that represent time-varying topics.
%
%
Ref.~\cite{Candecom} proposed  in $1970$ the Canonical Decomposition (CANDECOM), also called parallel factorization \\
(PARAFAC, \cite{harshman1970fpp}), which can be regarded as a generalization to tensors of singular value decomposition (SVD).
Maintaining the interpretability of the factors usually requires to achieve factorization under non-negativity constraints, leading to techniques such as non-negative matrix or tensor factorization (NMF and NTF).
Tensor factorization to detect latent structures has  been extensively used in several domains such as signal processing, psychometrics,  brain science, linguistics and chemometrics~\cite{Shashua:2005:NTF:1102351.1102451,Cichocki:1253894,VandeCruys:2009:NTF,Beyondstreams,Wang:2011}.


\vspace{0.5cm}
\section{Data and Representation}
\label{sec:data}
\subsubsection{Notation}
The following notations are used throughout the paper.
Scalars are denoted by lowercase letters, e.g., $x$, and vectors are denoted by boldface lowercase letters, e.g., $\textbf x$,
where the $i$-th entry is $x_i$.
Matrices are denoted by boldface capital letters, e.g., $\textbf X$, where the $i$-th column of matrix $\textbf X$ is $\textbf x_i$, and the $(i, j)$-th entry is $x_{ij}$.
Third order tensors are denoted by calligraphic letters, e.g., $\mathcal{A}$.
The $i$-th slice of $\mathcal{A}$, denoted by $\textbf{A}_i$,
is formed by setting the last mode of the third order tensor to $i$.
The $(i, j)$-th vector of $\mathcal{A}$, denoted by $\textbf{a}_{ij}$,
is formed by setting the second to last and last modes of $\mathcal{A}$ to $i$ and $j$ respectively,
and the $(i,j,k)$-th entry of $\mathcal{A}$ is $a_{ijk}$.

\subsubsection{Twitter Dataset}
The Emoto dataset consists of around 14 million tweets collected during the London 2012 Summer Olympics using the public Twitter Streaming API.
All tweets have at least one of 400 keywords,
including common words used in the Olympic Games -- like athlete, olympic,
sports names and twitter accounts of high followed athletes and media companies.
Tweets were collected during all the interval of 17 days comprising the Olympic Games, from July 27 to August 12 2012.

\subsubsection{Event Schedule}
In order to investigate the relation between the extracted time-varying topics and the sport events of the Olympic Games, we use the schedule available on the official London 2012 Olympics page\footnote{http://www.london2012.com/schedule-and-results/}, where the starting time and duration of most events is reported together with metadata about the type of event (discipline, involved teams or countries, etc.)

\subsubsection{Data Preprocessing }
For the text analysis performed in this paper, URLs are removed from the original tweet content.
The remaining text is used to build a vocabulary composed of the most common 30,000 terms,
where each term can be a single word, a digram or a trigram.
352 common words of the English language are also removed from the vocabulary.

In order to localize Twitter users, we examine the user profile descriptions
and use an adapted version of GeoDict\footnote{https://github.com/petewarden/geodict} to identify, if possible, the user country.
To study the relation between the extracted topical activity and the schedule of the Olympic events,
we focus on tweets posted by users located in the UK, only. This allows us to avoid potential confusion
arising from tweets posted in countries, such as the USA, where Olympics events were broadcasted with delays
of several hours due to time zone differences. This selection leaves us with a still substantial amount of data
(about one third of the full dataset) and simplifies the subsequent temporal analysis, even though it probably oversamples
the attention payed to events that involved UK athletes. 

For the scope of this study, we represent the 
data as a sparse third-order tensor $\mathcal T \in \mathbb{R}^{I \times J \times K}$, with $I$ tweets, $J$ terms and $K$ time intervals.
We aggregate the tweets over 1-hour intervals, for a total of $K=408$ intervals. 
The tensor $\mathcal T$ is sparse: the average number of terms (also referred as features in the following) for each tweet is typically no more than 10, compared to the 30k terms of our term vocabulary.
Moreover, as each tweet is emitted at a given time, each interval $k$ has a limited number of active tweets, $I_k$.
A tensor slice $\textbf{T}_k \in \mathbb{R}^{I \times J}$ is a sparse matrix with non-zero values only for $I_k$ rows.
$\textbf{T}_k$ represent the sparse tweet-term matrix observed at time $k$. The term values ${t}_{ijk}$ for each tweet $i$ are normalized using the standard Term Frequency and Inverse-Document Frequency (TF-IDF) weighting,
${t}_{ijk} = \mathrm{tf}(i, j) \times \mathrm{idf}(j)$,
where $\mathrm{tf}(i, j)$ is the frequency of term $j$ in tweet $i$, and 
$
\mathrm{idf}(j) = \log{\frac{\left|D\right|}{1+\left|\{d : j \in d\}\right|}}
$
where $|D|$ is the total number of tweets and $\left|\{d : j \in d\}\right|$ is the number of tweets where the term $j$ appears.

\subsubsection{Visualizing Topics over Time}
The methods that we present in this paper are able to extract topical-temporal structures from $\mathcal T$.
Such topical-temporal structures can be represented a stream matrix $\textbf{S} \in \mathbb{R}^{R \times K}$ with $R$ topics and $K$ intervals.
Each component $R$ is also characterized by a term-vector $\textbf{h} \in \mathbb{R}^{J}$ that defines the most representative terms for that component.
In order to visualize such topical-temporal structures represented as a stream matrix,
we use the method described by Byron and Wattenberg~\cite{byron2008stacked},
which yields a layered stream-graph visualization. 

\section{Masked Non-negative Tensor Factorization}
\label{sec:TF}


\subsubsection{Problem Statement}

As explained in section \ref{sec:data}, the tensor $\mathcal T \in \mathbb{R}^{I \times J \times K}$ with $I$ tweets, $J$ terms 
and $K$ intervals is a natural way to represent the tweets and their contents with respect to the time.
 The tensor has the advantage to  directly encompass
 the relationship between tweets posted at different hours and consequently between topics of the different hours. The tensor factorization as described below allows to uncover topics together with their temporal pattern.

\begin{figure}[!htbp]
\begin{center}
\includegraphics[width=9cm]{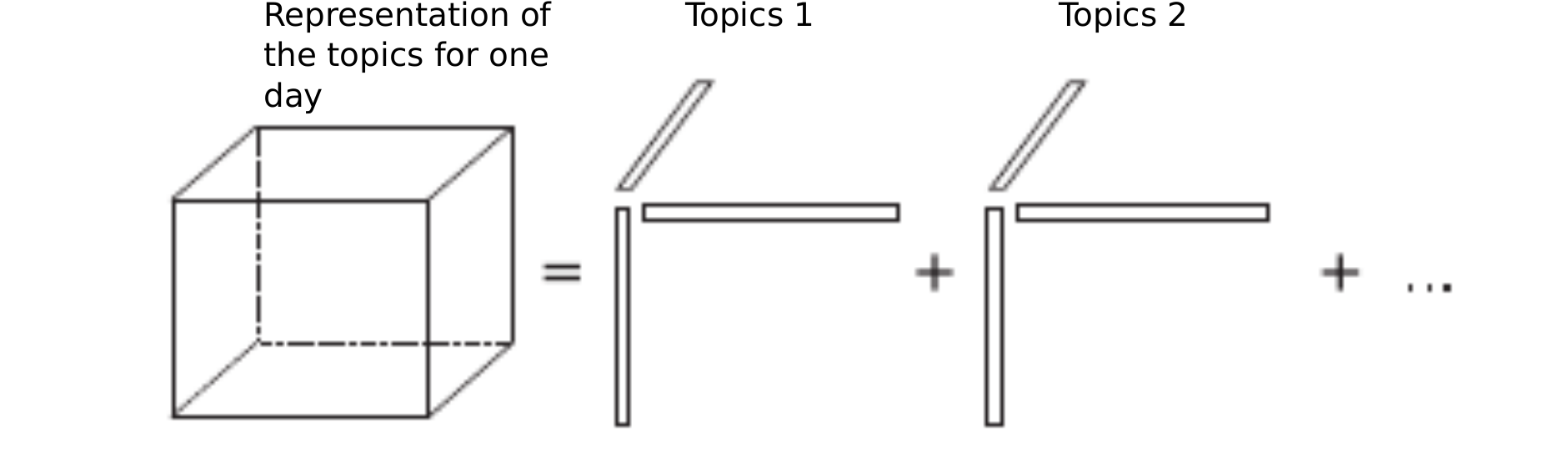}
\caption{Representation of a Kruskal decomposition. The cube corresponds to the tensor to be factorized while the rectangles represent the vectors. In the Twitter case, each of the rank-one tensor would correspond to the description of one topics.}
 \label{drawing_factorization}
\end{center}
\end{figure}

 Before describing the process of factorization itself and its output, one needs to introduce the concept of canonical decomposition (CP). CP in $3$ dimensions aims at 
 writing a tensor $\mathcal T \in \mathbb{R}^{I \times J \times K}$
 in a factorized way that is the sum of the outer product of three vectors:
\begin{equation}
\mathcal{T} = \sum_{r=1}^{R_\mathcal{T}} \,  \mathbf{a}_r \circ \mathbf{b_r} \circ \mathbf{c_r}
\label{approx}
\end{equation}
where the smallest  value of $R_\mathcal{T}$ for which this relation exists, is the rank of the tensor $\mathcal{T}$. In other words, the tensor $\mathcal{T}$ is expressed with a sum of rank-$1$ tensors.
The set of vectors
$a_{\{1,2,\dots,R\}}$ (resp. $b_{\{1,2,\dots,R_\mathcal{T}\}}$,$c_{\{1,2,\dots,R_\mathcal{T}}\}$) can be re-written as a matrix $\mathbf{A} \in \mathbb{R}^{I \times R_\mathcal{T}}$ 
(resp $\mathbf{B} \in \mathbb{R}^{J \times R_\mathcal{T}} $,$\mathbf{C} \in \mathbb{R}^{K \times R_\mathcal{T}}$) where each of the $R_\mathcal{T}$ vectors
is a column of the matrix. The decomposition of Eq.~\ref{approx} can also be represented in terms of the three matrices $\mathbf{A},\mathbf{B},\mathbf{C}$ as  $\llbracket \mathbf{A},\mathbf{B},\mathbf{C} \rrbracket$.
A visual representation of such a factorization, also called Kruskal decomposition,
is displayed on Fig.~\ref{drawing_factorization}.

\subsubsection{Factorization Methodology}

Regarding the extraction of topics, the aim is not to decompose the tensor in its exact form but to approximate the tensor by a sum of rank-$1$ tensors with a number of terms smaller than the rank of the original tensor.
This number $R$ corresponds to the number of topics that we want to extract (see Fig.~\ref{drawing_factorization}).
Such an approximation of the tensor leads to minimize the difference between $\mathcal{T}$ and $\llbracket \mathbf{A},\mathbf{B},\mathbf{C} \rrbracket$:
\begin{equation}
\min\limits_{\mathbf{A},\mathbf{B},\mathbf{C}} {\Vert \mathcal{T} -\llbracket \mathbf{A},\mathbf{B},\mathbf{C} \rrbracket \Vert}^2_F
\label{3d-problem}
\end{equation}
where $\Vert  \Vert $ is the Frobenius norm.
We transform the $3$-dimensional problem (Eq.~\ref{3d-problem}) in $2$-dimensional sub-problems by unfolding the tensor $\mathcal{T}$ in three different ways. This process called matricization
gives rise to three modes $\mathbf{X}_{(1)},\mathbf{X}_{(2)},\mathbf{X}_{(3)}$. The mode-$n$ matricization consists of linearizing all the indices of the tensor except $n$. The three 
resulting matrices have respectively a size of $I \times JK$,$J \times IK$ and $K\times IJ$. Each element of the matrix $\mathbf{X}_{(i=1,2,3)}$ corresponds to one element of the tensor
 $\mathcal{T}$ such that each of the mode contains all the values of the tensor.
Due to matricization, the factorization problem given by Eq.\ref{approx} can be reframed in factorization of the three modes. In other terms, maximizing the likelihood between 
$\mathcal{T}$ and $\llbracket \mathbf{A},\mathbf{B},\mathbf{C} \rrbracket$ is equivalent to minimizing 
the difference between each of the mode and their respective approximation in terms of $\mathbf{A},\mathbf{B},\mathbf{C}$.
%
The factorization problem (PARAFAC) in Eq.\ref{3d-problem} is converted to the three following sub-problems where we added a condition of non-negativity of the three modes:
\begin{eqnarray}
\min\limits_{\mathbf{A}\geq 0} {\Vert \mathbf{X}_{(1)} -\mathbf{A}(\mathbf{C}\odot \mathbf{B})^T \Vert}^2_F\\
\min\limits_{\mathbf{B} \geq 0} {\Vert \mathbf{X}_{(2)} -\mathbf{B}(\mathbf{C}\odot \mathbf{A})^T \Vert}^2_F\\
\min\limits_{\mathbf{C} \geq 0} {\Vert \mathbf{X}_{(3)} -\mathbf{C}(\mathbf{B}\odot \mathbf{A})^T \Vert}^2_F
\label{2d-problem}
\end{eqnarray}
where $\odot$ is the Khatri-Rao product which is a columnwise Kronecker product, i.e. such that
$\mathbf{C}  \odot \mathbf{B}  =[c_1 \otimes b_1 c_2 \otimes b_2  \dots c_r \otimes b_r]$. If $\mathbf{C} \in \mathbb{R}^{K \times R}$ 
and $\mathbf{B} \in \mathbb{R}^{J \times R}$, then the Khatri-Rao product $\mathbf{C} \odot \mathbf{B} \in \mathbb{R}^{KJ \times R}$.
In our case of study, $\mathbf{A},\mathbf{B},\mathbf{C}$  will give each access to a different information: $\mathbf{A}$ allows to know at which topic belongs a tweet, $\mathbf{B}$ gives the definition of the topics
with respect to the features and $\mathbf{C}$ gives the temporal activity of each topic.

Several algorithms have been developped to tackle the PARAFAC decomposition. The two most common are one method based on the projected gradient and the Alternating Least square method (ALS). The first one is convenient for its ease of implementation
and is largely used in Singular Value Decomposition (SVD) but converges slowly. 
In the ALS method, the modes are deduced successively by solving Eq \ref{2d-problem}. In each iteration, for each of the sub-problem, two modes are kept fixed while the third one is computed. This process is repeated until convergence.
In our case, we use a nonnegativity constraint to make the factorization  better posed and the results meaningful.
 One thus uses nonnegative ALS (ANLS \cite{Paatero}) combined with a block-coordinate-descent method in order to reach the convergence faster.
Each of the step of the algorithm needs to take into account the Karush-Kuhn-Tucker (KKT) conditions to have  a stationary point. Our program is based on the algorithm implemented by \cite{Park-NTF}.

\subsubsection{Masked Adaptation of the NTF}

We cannot directly perform the NTF on the tensor [Tweets $\times$ Features $\times$ Interval] built as explained aboved as this tensor has a ``block-disjoint'' structure peculiar to the tweets. 
Indeed each tweet has non-zero values only at one interval because a tweet is emitted only at a given time. Each interval $k$ has only $I_k$ active tweets. In each slice $\textbf{T}_k$ of the tensor, only $I_k$ rows have meaningful values.
 So, we are only interested in reproducing the tensor part which contains the meaningful values. In order to focus on these meaningful values, one needs to consider an adapted version of the tensor $\mathcal{T}$. 
We first consider  the tensor $\mathcal{T}$ built as explained above. We generate a first set of matrices $\mathbf{A},\mathbf{B},\mathbf{C}$ which could approximate the tensor. At the next step, one tries to decompose a 
tensor $\bar{\mathcal{T}}$ where the values are a combination of the values of $\bar{\mathcal{T}}$  and of the values of $\llbracket \mathbf{A},\mathbf{B},\mathbf{C} \rrbracket$.  More exactly, this tensor has the same size than $\mathcal{T}$
and the same values than $\mathcal{T}$ for the rows $I_k$ of each slice $\mathbf{\bar{T}}_k$. The complementary values are given by $\llbracket \mathbf{A},\mathbf{B},\mathbf{C} \rrbracket$.
In other terms, at each step, the tensor that we approximate is updated by:
\begin{equation}
\bar{\mathcal{T}}= {\mathcal{T}} \boxdot {\mathcal{W}} + (1-{\mathcal{W}})\llbracket \mathbf{A},\mathbf{B},\mathbf{C} \rrbracket 
\end{equation}
where $\boxdot$ is the Hadamard product (element-wise product) and ${\mathcal{W}}$ is a binary tensor of the same size than ${\mathcal{T}}$ with $1$-values only when the values of ${\mathcal{T}}$ at this position are meaningful.
The particular structure of the tensor (disjoint blocks in time) could be perceived as a ``missing values'' problem in the tensor, this problem has been for example tackled in \cite{royer:hal-00725289}. 

%
Concretely, the implementation is an adaptation of a Matlab program \cite{Park-NTF}  which uses the Tensor Toolbox \cite{Kolda:2009:TDA:1655228.1655230}. This adaptation includes the introduction of a
mask (via the tensor of weight) as mentionned above and the rewriting of some operations to avoid memory issues. This point is not detailed here as it is not part of the main point of the paper.

\subsubsection{Stream Matrix Construction}

We calculate the strength of each topic with respect to the time by using both the information about the link between each topic and each tweet  and about temporal pattern of the topics. These informations are available through
$\mathbf{A}$ and $\mathbf{C}$  and the consequent strength of a topic $r$ on each interval of time $k$ is given by:
\begin{equation}
 s_{rk}=\sum_{i|k} a_{ir}*c_{kr}
\end{equation}
where $\sum_{i|k}$ is a sum over the tweets indexed by $i$ occurring at the interval indexed by $k$. The set of elements $s_{\{r,k\}}$ with $r=\llbracket1,R\rrbracket $ and $k=\llbracket 1,K \rrbracket$ forms the
stream matrix $\mathbf{S}$.
Each topic is then defined by a terms vector and each of this term vector is given by a column of $\mathbf{B}$.

\section{Agglomerative Non-negative Matrix Factorization}
\label{sec:MF}




\subsubsection{Non-negative Matrix Factorization}


For each tensor slice $\textbf{T}_k$, we compute a non-negative factorization by minimizing the following error function,
\begin{equation}
\min\limits_{W,H} {\Vert \textbf{T}_k - \textbf{W}^{(k)} \textbf{H}^{(k)} \Vert}^2_F \, ,
\label{nmf-minimization}
\end{equation}
where $\Vert  \Vert $ is the Frobenius norm, subject to the constraint that the values in $\textbf{W}^{(k)}$ and $\textbf{H}^{(k)}$ must be non-negative.
The non-negative factorization is achieved using the projected gradient method with sparseness constraints, as described in~\cite{lin2007projected,hoyer2004non}.
The factorization produces
a matrix of left vectors $\textbf{W}^{(k)} \in \mathbb{R}^{I_k \times F}$
and a matrix of right vectors $\textbf{H}^{(k)} \in \mathbb{R}^{F \times J}$, where $F$ is the number of components used in the decomposition.
The matrix $\textbf{H}^{(k)}$ stores the term vectors of the extracted components at interval $t$.
The matrix $\textbf{W}^{(k)}$ is used to calculate the 
strength of each extracted component, which are represented in a matrix $\textbf{Z} \in \mathbb{R}^{F \times K}$ given by
\begin{equation}
z_{fk} = \sum\limits_{i=1}^{I_k} { \frac{ w^{(k)}_{if} } {    \sum\limits_{f'=1}^F  {w^{(k)}_{if'} }     } }
\label{weight_matrix}
\end{equation}
where $ z_{fk}$ is the strength of factor $f$ at interval $k$.

\subsubsection{Component Clustering}

In order to track topics over time, we need to merge components into topics depending on how similar they are.
Since each component is defined by a term vector,
we can calculate a similarity matrix of all possible pairs of term vectors using cosine similarity.
This matrix is fed to a standard agglomerative hierarchical clustering algorithm,
known as UPGMA~\cite{Sokal1958},
that at each step 
combines the two most similar clusters into a higher-level cluster.
Cluster similarity is defined in terms of average linkage:
that is, the distance between two clusters $c_1$ and $c_2$ is defined as the average
of all pair-wise distances between the children of $c_1$ and those of $c_2$.




The hierarchical clustering produces a tree that can be cut at a given depth to yield a clustering at a chosen level of detail.
That is, by varying the threshold similarity we use for the cut we can go from a coarse-grained topical structure, with few clusters that may merge unrelated topics, to a fine-grained topical structure, with many clusters that may separate term vectors that otherwise could be regarded as the same continuous topic over time.
The cut threshold needs to be chosen based on criteria that depends on the application at hand.


Each choice for the cut yields a number of clusters $C$ and a 
map function $\mathcal{C}(r,f) \rightarrow k$ that associates the component index $f$ at time interval $k$
to a topic cluster $r$.
This function collects all components associated to cluster $r$ in a set $\mathbb{C}_{rk}$ for each interval $k$.

\subsubsection{Stream Matrix Construction}

When constructing the stream matrix, the number of topics $R$ in the stream matrix is given by the number of clusters generated by the clustering step.
%
In order to calculate the entries $s_{rk}$ of the resulting stream matrix $\textbf{S}$, we aggregate the strengths of the clustered components.
%
%
We build a stream matrix $\textbf{S} \in \mathbb{R}^{R \times K}$, with $R$ topics and $K$ intervals, given by


\begin{equation}
  {s_{rk}} = {\sum\limits_{f \in \mathbb{C}_{rk}} { z_{fk} }}
\end{equation}

Finally, we extract the term vectors that are associated to each cluster.
Each cluster will be associated to a term vector $\textbf{h}^{(k)}_{r} \in \mathbb{R}^{J}$ that is the average of all term vectors $\textbf{h}^{(k)}_{f}$ associated to that cluster in the component clustering step.




\vspace{0.5cm}
\section{Analysis of the Olympics Dataset}
\label{sec:results}
We now move to the analysis of the London 2012 Twitter dataset and its relation with the known schedule of the Games. We focus on one representative day, July 29th, during which several sport events took place at different times and concurrently. We use both topic detection methods, show the signals they extract, and check to what extent they are capable of extracting signals that we can understand in terms of the schedule.

The topic detection methods are set up as follows. For the masked NTF method, we decompose the tensor using a fixed number of components, using a tolerance value of $10^{-4}$ for the stopping condition, and limiting the number of iterations to $50$.
For the agglomerative NMF method, we decompose each interval matrix using a fixed number of components. We use a tolerance value of $10^{-4}$ for the stopping condition, and limit the number of iterations to $20$.
We use 250 topics for the Masked NTF, and 50 components per time intervals in the Agglomerative NMF.


%
\begin{figure*}[!ht]
\begin{center}
 \includegraphics[width=1.7\columnwidth]{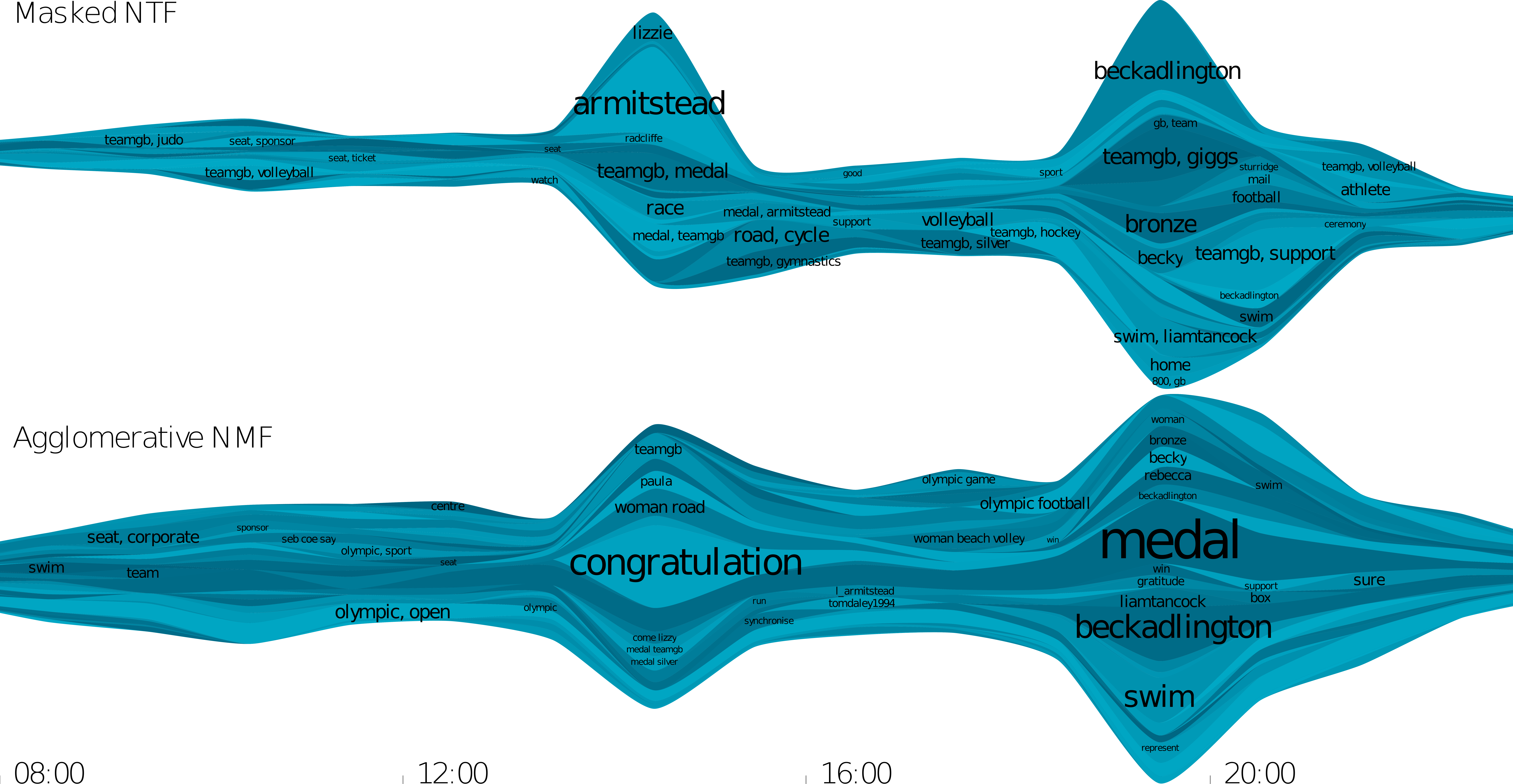}
\caption{
Streamgraph visualization of the stream matrices generated by the Masked NTF (top) using 50 topics, and by the Agglomerative NMF (bottom) using 20 components per interval and a total of 150 clusters.
Interactive streamgraph visualizations for a few use cases are available at 
\href{http://www.datainterfaces.org/2013/06/twitter-topic-explorer/}{http://www.datainterfaces.org/2013/06/twitter-topic-explorer/}. 
\label{bigstream2}
}
\end{center}
\end{figure*}

Figure~\ref{bigstream2} shows a streamgraph representation of time-varying topics extracted using the two methods we have discussed.
Two global activity peaks are visible in both streamgraphs: the peak at about 2.30pm UTC
was triggered when Elizabeth Armistead won the silver medal in road cycle;
the peat at about 7pm UTC is driven by the bronze medal in 400m freestyle to Rebecca Adlington.
In the stream graphs, for clarity, each topic is annotated using only its topmost weighted term.
This makes it difficult to assess a visual correspondence between the same topics across the two representations,
as the term with top weight may be different for the two term vectors even though the vectors are overall very similar
(in terms of cosine similarity).
On closer inspection, many precise correspondences can be established between the topics extracted
by the Masked NTF method and those extracted by the Agglomerative NMF method:
for example, the topic \textit{armistead} in the top streamgraph matches the topic \textit{congratulation} in the bottom one.
An interactive streamgraph visualization of the London 2012 Twitter dataset
is available at
\href{http://www.datainterfaces.org/projects/emoto/}{http://www.datainterfaces.org/projects/emoto/}.


%
\subsection{Comparison with the Olympics Schedule}

\subsubsection{Event Selection}
In order to show the possible correspondence between the extracted topics and sport events,
we manually annotate the schedule collected from the official London 2012 Olympics page for July 29th, 2012.
As the number of events in a day can be substantial and we want to focus on events with higher impact on social media, we retain events that are either finals or team sports match.
We annotate each event with a set of at most three terms extracted from the schedule, as described in Section \ref{sec:data}.
For a team sport, we use the sport name and the countries of the two teams, otherwise, we put the name of the sport and its characteristics, e.g., the discipline for swimming.

\subsubsection{Matching Topics and Events}
For each event, we use a matching criteria to select one of the extracted topics from each of the set of topics produced by the methods.
Since we want to select a topic in which all event annotated terms appear with a high weight in its term vectors, 
we define our matching score based on the geometric average of the weights of the event annotated terms in the topic's term vectors:
\begin{equation}
\langle  w \rangle = \sqrt[n]{h_{w_1 r} \, h_{w_2 r} \, \dots \, h_{w_n r}}
\end{equation}
For Masked NTF, for each event, we choose the topic with the highest corresponding geometric average $\langle  w \rangle$.
In the agglomerative NMF case, for each event, we choose the topic with the highest corresponding geometric
average $\langle  w \rangle$ weighted by $\log(n)$ where $n$ is the number of components in the selected cluster.
We use $\log(n)$ in order to favor the selection of clusters with a higher number of aggregated components,
otherwise the most detailed clusters which aggregates only one component are always selected.
Since the Agglomerative NMF method produces a tree structure in which each node agglomerates a set of components and represents topic activity,
we have to calculate such matching result for each node, and select the node for which such matching result is the highest.



\subsubsection{Results and Observations}

\begin{figure*}[!htbp]
\centering
\subfigure[Masked NTF]{
\label{fig:distht09}
	\includegraphics[width=0.47\textwidth, keepaspectratio]{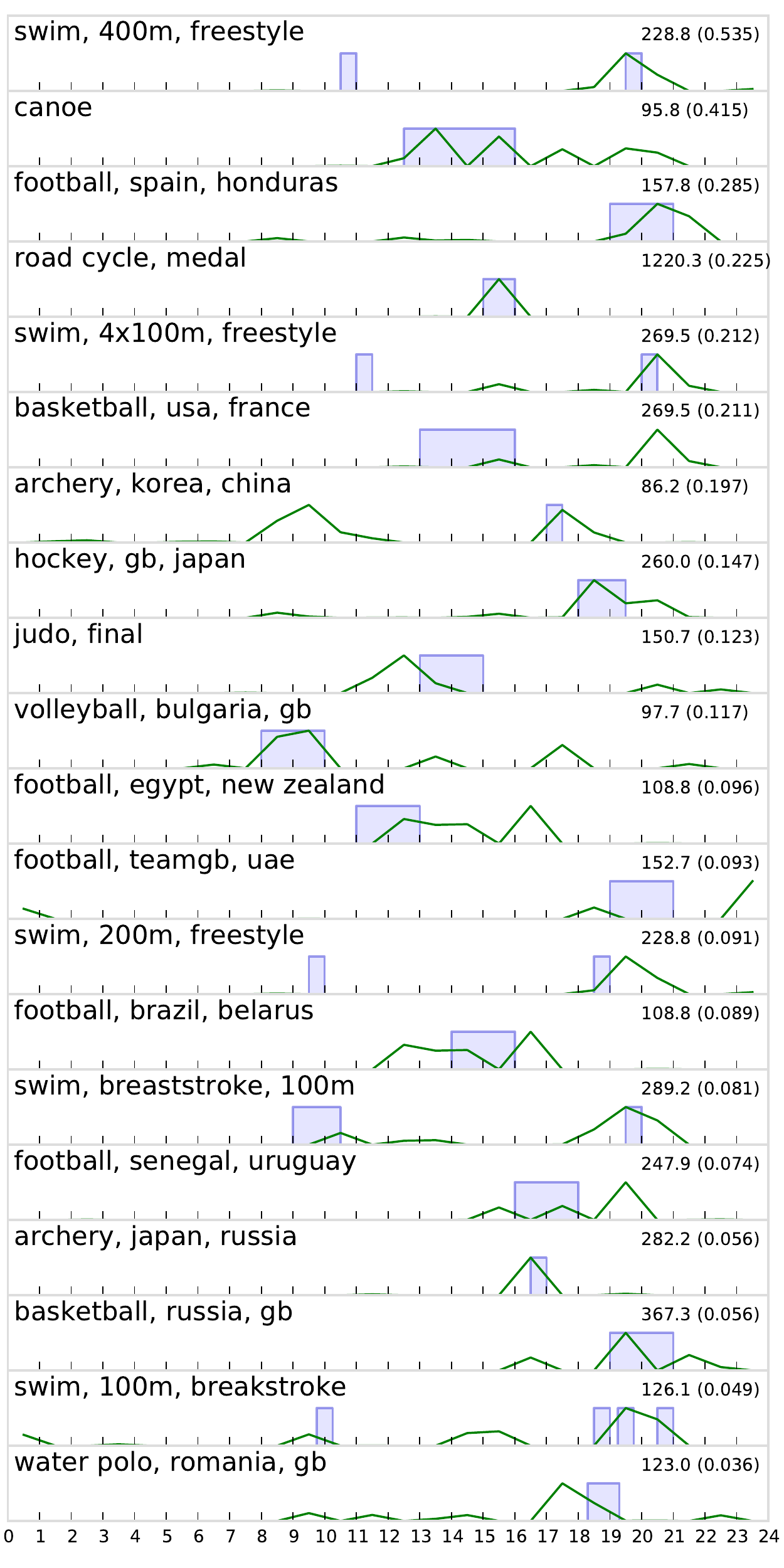}

}
\subfigure[Agglomerative NMF]{
	\label{fig:distobg}
	\includegraphics[width=0.47\textwidth, keepaspectratio]{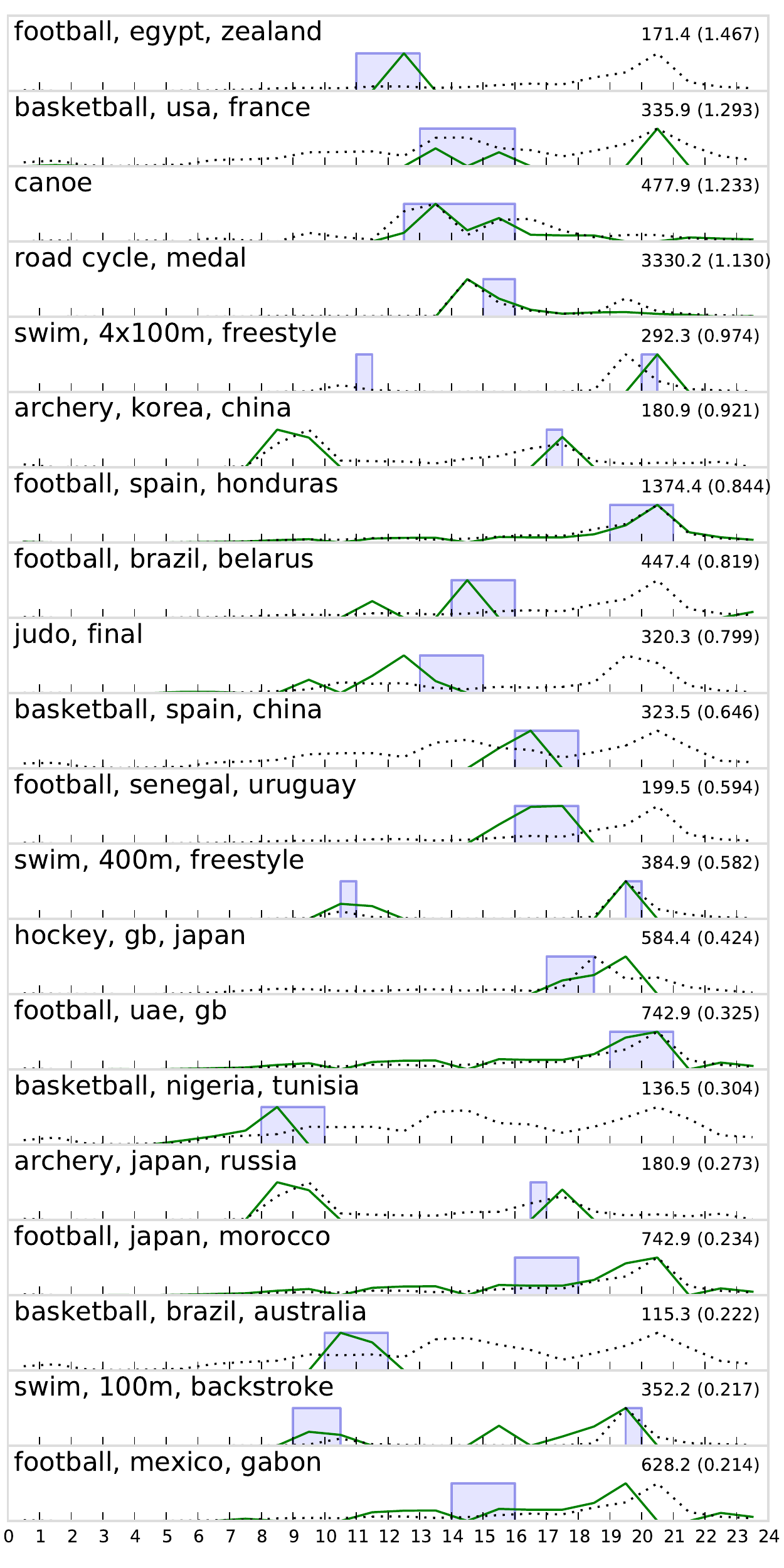}
}
\caption[]{
The top $20$ most representative schedule events regarding the matching weight of its annotations with the term vectors of an extracted topic.
In the left, for each event, we show the topic extracted by the Masked NTF method for which the matching weight is the highest, and in the right we show the topic extracted by the Agglomerative NMF method for which the matching weight is the highest.
Since we are showing the topmost $20$ schedule events regarding the matching weight, the events are sorted by such matching weight.
%
%
On the top left corner of each event, we show its annotated terms, along with the exact interval in which the event happened according to the official Olympics schedule (shaded blue area).
The solid green line shows the temporal structure of the topic with higher matching weight along the 24 hours of July 29.
The values in the top right side shows the value for the peak of the temporal structure, which roughly represents the amount of activity for such topic,
and, between parenthesis, the matching weight for the given topic.
In the Agglomerative NMF side (on the right) we show as a dotted line the activity in time for the given terms regarding the number of tweets that have such terms (tweet count).
%
%
%
%
%
%
%
}
\label{fig:distinf}
\end{figure*}

At this point, we have, for each event, a topic which was selected in each method, and the corresponding matching result.
In Figure~\ref{fig:distinf}, we show the schedule events for the top $20$ highest matching results.
In the lefthand figure, we show, for each one of the top $20$ matching results, the topic extracted by the Masked NTF method,
while in the righthand figure, we show the topic extracted by the Agglomerative NMF method.
The results are sorted by the corresponding matching weight.

For each event, on its top left corner, we show the manually annotated terms used for the matching.
The shaded blue area shows the exact interval during which the event was occurring according to the official Olympics schedule.
In the same area, the solid green line represents the temporal structure of the topic with higher matching result according to our matching criteria.
Such values roughly represent the amount of activity for such topic and are normalized according to the peak of activity.
We show the value for this peak in the top right side, along with the matching results between parenthesis.
In the Agglomerative NMF graph (on the right) we show as a dotted line the activity in time for the given terms regarding the number of tweets that have such terms (tweet count).
We remark that by considering only the dotted line the timing of many events on the right side of the figure does not match the schedule timings, i.e., merely counting tweets is not sufficient at this resolution level.
We also measured the number of tweets where the terms are co-ocurring, and in this case the number of tweets is so small that it does not allow  the detection of any structure in time.

\begin{figure*}[!htbp]
\begin{center}
 \includegraphics[width=1.8\columnwidth]{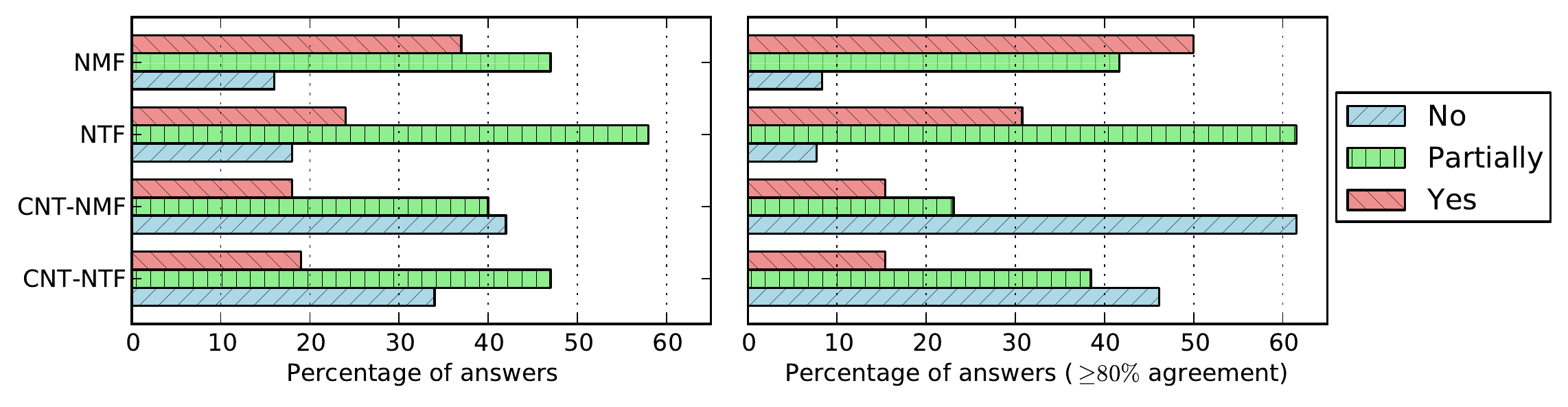}
\caption{
Crowdsourced evaluation of the topical activity profiles for selected sport events (see main text) of the London 2012 Olympics dataset obtained by using the different topic detection methods.
\label{crowdsourced}
}
\end{center}
\end{figure*}

We evaluated these activity profiles using the CrowdFlower Web-based crowdsourcing platform (restricted to Amazon Mechanical Turk workers).
Each work unit asks a worker to visually inspect and compare two timelines: the one to be evaluated, and a reference timeline corresponding to the known time intervals for sport events taken from the Olympic schedule.
Each work unit looks like a row from Figure~\ref{fig:distinf}.
Our evaluation was based on 100 work units evenly distributed among 5 types:
1) (NMF) work units based on the results of Agglomerative NMF; 2) (CNT-NMF) work units with activity profiles generated by simply counting the number of tweets with the terms used in matching the NMF topics; 3) (NTF) work units from the Masked NTF approach; 4) (CNT-NMF) same as (CNT-NMF) for Masked NTF; 5) synthetic work units (``gold'' units) used to assess worker quality.
For each work unit, we asked the workers whether the two timelines matched exactly (Yes), matched partially (Partially) or not at all (No).
95\% of the judgments for gold work units were correct. We only retained those users who correctly judged more than 80\% of the gold units.
%
Figure~\ref{crowdsourced} shows the distribution of judgements for the different types of work units.
The left hand side of the figure shows the distribution obtained for all work units,
while the right hand side shows the distribution restricted to work units with more than 80\% of agreement across different workers.
According to this evaluation, both NTF and NMF outperform the count-based methods.

We see that for most of the events there is a close temporal alignment between the event schedule and the topic structure, at the scale of the hour or less.
We see that such temporal alignment is much closer than when compared to the peaks of activity generated by counting tweets.

We observe that the mismatches in the temporal alignment are caused by two different factors.
The first one is due to a low matching results, like the event annotated with (football, mexico, gabon).
It means that the term vectors for the given topic does not represent with high confidence the terms used to annotate the event.
The second one is due to a different behaviour in collective attention.
This happens for example in the case of swimming events, where the first part of the event is related to eliminatories and
the second part is related to the finals.
In such cases, the peak in activity arrives when the event finishes and the attention goes to the winner.



%

%
%
%
%
%
%

\section{Summary and Future Work}
\label{sec:conclusions}
The topic detection techniques we discussed here afford tracking the attention that a community of users devotes to multiple concurrent topics over time, teasing apart social signals that cannot be disentangled by simply measuring frequencies of term or hashtags occurrences. This allows to capture the emergence of topics and to track their popularity with a high temporal resolution and a controllable semantic granularity.
The comparison with an independently available schedule of real-world events shows that the response of Twitter to external driving retains a great deal of temporal and topical information about the event schedule, pointing to more sophisticated uses of Twitter as a social sensor.

The work described here can be extended along several directions.
It would be interesting to develop and characterize on-line versions of the techniques we used here,
so that topic emergence and trend detection could be carried out on live microblog streams.
Because of its temporal segmentation, the Agglomerative NMF case lends itself rather well
to on-line incremental computation, whereas a dynamic version of the Masked NTF
technique would be more challenging to achieve.

Another interesting direction for future research would be to augment the tweet-term-time tensor with a fourth dimension representing the location of the users, so that the latent signals we extract could expose correlation between topics, time intervals and locations, exposing geographical patterns of collective attention and their relation to delays, e.g., in the seeding by mass media across different countries.




\section*{Acknowledgements}
\label{sec:acks}
\small
The Authors acknowledge the Emoto project \href{www.emoto2012.org}{www.emoto2012.org} and its partners for access to the Twitter dataset on the London Olympics 2012.
The Authors acknowledge inspiring discussions with Moritz Stefaner and Bruno Goncalves.
The Authors aknowledge partial support from the Lagrange Project of the ISI Foundation funded by the CRT Foundation, from the Q-ARACNE project funded by the Fondazione Compagnia di San Paolo, and from the FET Multiplex Project (EU-FET-317532) funded by the European Commission.

\small
\bibliographystyle{abbrv}

\end{document}